\documentstyle[12pt]{article}

\newlength{\dinwidth}
\newlength{\dinmargin}
\newcommand{\resection}[1]{\setcounter{equation}{0}\section{#1}}

\begin{document}
\vspace*{7cm}
\begin{center}
  \begin{Large}
  \begin{bf}
PSEUDO GOLDSTONES AT FUTURE COLLIDERS FROM THE EXTENDED BESS MODEL$^*$\\
  \end{bf}
  \end{Large}
  \vspace{10mm}
  \begin{large}
R. Casalbuoni, S. De Curtis and D. Dominici\\
  \end{large}
Dipartimento di Fisica, Univ. di Firenze\\
I.N.F.N., Sezione di Firenze\\
  \vspace{5mm}
  \begin{large}
P. Chiappetta\\
  \end{large}
Centre Physique Th\'eorique, CNRS Luminy, Marseille\\
  \vspace{5mm}
  \begin{large}
A. Deandrea and R. Gatto\\
  \end{large}
D\'epartement de Physique Th\'eorique, Univ. de Gen\`eve\\
  \vspace{5mm}
\end{center}
  \vspace{2cm}
\begin{center}
UGVA-DPT 1994/03-845\\\
hep-ph/9403305\\\
March 1994
\end{center}
\vspace{1cm}
\noindent
$^*$ Partially supported by the Swiss National Foundation
\newpage
\thispagestyle{empty}
\begin{quotation}
\vspace*{5cm}
\begin{center}
  \begin{bf}
  ABSTRACT
  \end{bf}
\end{center}
  \vspace{5mm}
\noindent
We consider the production of the lightest pseudo-Goldstone bosons at
future colliders through the vector resonances predicted by the extended BESS
model, which consists of an effective lagrangian parametrization with
dynamical symmetry breaking, describing scalar, vector and axial-vector bound
states in a rather general framework. We find that the detection of
pseudo-Goldstone pairs at LHC requires a careful evaluation of backgrounds.
For $e^+e^-$ collisions in the TeV range the backgrounds can be easily reduced
and the detection of pseudo-Goldstone pairs is generally easier.
\end{quotation}
\newpage
\setcounter{page}{1}
\def\lq{\left [}
\def\rq{\right ]}
\def\qq{Q^2}
\def\dmu{\partial_{\mu}}
\def\dmus{\partial^{\mu}}
\def\AA{{\cal A}}
\def\BB{{\cal B}}
\def\Tr{{\rm Tr}}
\def\gp{g'}
\def\gs{g''}
\def\ggs{\frac{g}{\gs}}
\def\mpp{M_{P^+}}
\def\mpm{M_{P^-}}
\def\mpt{M_{P^3}}
\def\mpz{M_{P^0}}
\def\eps{{\epsilon}}
\newcommand{\be}{\begin{equation}}
\newcommand{\ee}{\end{equation}}
\newcommand{\bea}{\begin{eqnarray}}
\newcommand{\eea}{\end{eqnarray}}
\newcommand{\nn}{\nonumber}
\newcommand{\dd}{\displaystyle}

\resection{Introduction}

The possibility of a new strong-interaction sector as responsible for
electroweak symmetry breaking, although difficult to formulate in a
quantitatively comprehensive scheme, is still considered as a possible
alternative to the theoretically unsatisfactory present formulation in terms of
elementary scalars. The earliest suggestion in this sense was technicolor \cite
{techno1}. The one-family technicolor model \cite{techno2} is based on four
techniquark doublets, 3 with colours plus one leptonic, and has thus a flavour
symmetry $SU(8)\otimes SU(8)\otimes U(1)$. Anomaly cancellation occurs
similarly as for the ordinary quark-lepton families. Condensation of
technifermions brings down to the diagonal flavour $SU(8)\otimes
U(1)$ group, giving a total of 63 Goldstone bosons. Three of
them provide for the longitudinal degrees of freedom of $W$ and $Z$. While
ordinary fermions remain massless at this stage, an extension of the theory,
called extended technicolor, generates fermion masses. At the same time however
it leads to difficulties related to the experimental limits on flavour-changing
neutral currents. Recently, thoughts have been devoted to this difficult
problem, with proposals referred to as walking technicolor \cite {techno3}.

Theories of dynamical symmetry breaking, avoiding the introduction of
fundamental scalar fields, generally lead to the prediction of
pseudo-Goldstone bosons, due to the breaking of a
large initial global symmetry group $G$. We have considered the production of
the lightest pseudo-Goldstone bosons (PGB's) at future colliders
through the vector resonances predicted by the extended BESS model
with $SU(8)\otimes SU(8)$ symmetry. This model \cite {bessu8} consists of an
effective lagrangian parametrization which describes scalar, vector and
axial-vector bound states in quite a general framework. For example, the
effective low energy phenomenology of ordinary technicolor would correspond
to a specialization of the extended BESS picture.

In the absence of a specific definite theory of the strong electroweak sector,
one would like to remain as general as possible, avoiding
specific dynamical assumptions. The BESS model (BESS = breaking electroweak
symmetry strongly) was essentially developed to provide for such a general
frame \cite{BESS}. The basic ingredients for the construction were custodial
symmetry and gauge invariance.

The original BESS was based on a minimal chiral structure $G=SU(2)_L\otimes
SU(2)_R$. To discuss physics related to possible pseudo-Goldstones one has to
use an extension of the model to a larger $SU(N)_L\otimes SU(N)_R$ chiral
structure. The extended BESS will contain explicit vector and axial vector
resonances and a number of pseudo-Goldstones. Through their mixing with $W$,
$Z$ and gluons, some of the spin-one resonances will couple to quarks and
leptons, and thus they will be produced at proton-proton and electron-positron
colliders at high energy.

Extended BESS can be taken for $N=8$, and, more particularly, specialized to
reproduce the low energy phenomenology of the ``historical'' $N=8$ technicolor.
The main new features brought by extended BESS into lower energy
phenomenology are a number of low mass pseudo-Goldstones and the appearance of
an additional singlet vector resonance, in addition to the vector triplet of
vector resonances typical of the original BESS model.

We recall that, in dynamical schemes for electroweak symmetry breaking, an
initial global invariance group $G$ is subsequently spontaneously broken into a
subgroup $H$ by the symmetry breaking dynamical mechanism. As long as
additional interactions (such as gauge interactions and others), which break
$G$, are neglected, the Goldstone bosons, which correspond to the generators
belonging to the quotient $G/H$, remain exactly massless. Among those
interactions are the standard model gauge interactions of the local group
$SU(3)\otimes SU(2)_L\otimes U(1)_Y$, which in general break the symmetry $G$.
The resulting effective interactions will break the degeneracy among the
initial vacua, or, saying it differently, induce an orientation within such
vacua.

The initially massless Goldstone bosons, which correspond to oscillations along
the directions connecting different vacua, will not remain all massless, due to
the induced vacuum orientation. Such massive scalars are the pseudo-Goldstone
bosons (PGB) (see Weinberg in \cite{techno1}). Among the interactions
responsible for the
induction of the vacuum orientation are the local gauge interaction of the
color electroweak group. They contribute to the pseudo-Goldstone masses, and
studies of their properties, within a class of models, have been carried out
\cite{massepeskin}.

However, in any dynamical symmetry scheme, this will not be the only source of
pseudo-Goldstone masses. In fact suitable terms must be present, responsible
for
the masses of quarks and leptons themselves. For instance, in extended
technicolor, one introduces gauge interactions which connect ordinary fermions
to technifermions. The chiral symmetry $G$, which is only related to the
technifermion sector, is broken, and one expects this to be a source of
pseudo-Goldstone masses.

The fact that these interactions are essentially the source of the
fermion masses leads one to expect that the induced pseudo-Goldstone masses
from
those interactions could in some way be related to the fermion masses. This
point was quantitatively examined \cite{masse} and we shall refer to such a
study for a choice of the pseudo-Goldstone mass spectrum adopted in the
present work.

\resection{PGB's at future colliders}

As we have mentioned in the introduction, for quantitative estimates of the
pseudo-Goldstone production cross-sections, we shall employ the
$SU(8)\otimes SU(8)$ extended BESS model. For earlier studies on PGB
phenomenology in technicolor theories we refer the reader to \cite{PGB} and
\cite{Lubicz} and references therein. We will denote the $SU(8)$ gauge fields
as $V^A=(V^a,{\tilde V}^a,V_D,V_8^\alpha,V_8^{a \alpha},V_3^{\mu i},
{\bar V}_3^{\mu i})$, where $\mu=(0,a)$ ($a$ being an $SU(2)$ index),
and $i=1,2,3$ is a color index. An analogous notation will be used for
the axial particles $A^A$ and  the Goldstone bosons $\pi^A$. The $SU(8)$
generators can be found in Appendix A of \cite {bessu8}. We shall use through
out this work the notations of ref. \cite{bessu8} and \cite{masse}.

The production is induced by the processes
\be f^++f^-\rightarrow \gamma,Z, V^3\rightarrow P^+ P^-
\ee
and
\be f_1+f_2\rightarrow W^\pm , V^\pm\rightarrow P^\pm P^0
\ee
where $P^\pm (P^0)$ denote the lightest charged (neutral) PGB's and $f$ denotes
a light fermion. The gauge bosons $V$ appearing in (2.1) (2.2) are the $V^a$.

The main point of the calculations performed in ref. \cite{masse} was to work
within the low energy effective theory, as characterized by the initial chiral
group $G$, its unbroken subgroup $H$, and the color electroweak group, assuming
that the information for the fermion mass mechanism can be embodied, from the
viewpoint of the low energy theory, into effective Yukawa couplings between
ordinary fermions and pseudo-Goldstones.

Most of the allowed Yukawa coupling constants, for those couplings which are
invariant under the color-electroweak group, are then related, within the low
energy expansion, to the fermion masses. The pseudo-Goldstone mass spectrum can
then be derived from the one-loop effective potential, which includes, besides
the ordinary gauge interactions, also the Yukawa couplings. The resulting
masses
are expected in general to lie in a natural range depending on the masses of
the heaviest fermions, that is the top and bottom quarks. In particular, those
states, which would remain massless in absence of the Yukawa couplings, are
expected to lie in a range situated around such heaviest fermions.

For the calculations performed in this work we shall adopt a possible PGB
spectrum obtained in ref.\cite{masse}. The PGB's have the following masses,
for the choice of parameters: $\Lambda=2~TeV$, $\alpha_s=0.12$ and
$m_t=150~GeV$:
\bea
& &M^{2}(\pi^{a})=0 \hskip 1.5 truecm a=1,2,3 \nn\\
& &M^{2}({\tilde \pi}^{\pm})={\Lambda^{2}\over \pi^{2}
v^{2}}\left( m_t^2+m_b^2 \right)=(388~GeV)^2 \nn\\
& &M^{2}\left( {{\tilde\pi}^{3}-\pi_{D}\over \sqrt{2}}\right)={2 \Lambda^{2}
\over \pi^{2} v^{2}}m_b^2 =(18~GeV)^2 \nn\\
& &M^{2}\left( {{\tilde\pi}^{3}+\pi_{D}\over \sqrt{2}}\right)={2 \Lambda^{2}
\over \pi^{2} v^{2}}m_t^2  =(548~GeV)^2\nn\\
& &M^{2}(\pi_{8}^{\alpha \pm})={\Lambda^{2} \over
4\pi^{2} v^{2}}\left[m_t^2+m_b^2+{9\over 2}v^2 g_s^2 \right] =(952~GeV)^2 \nn\\
& &M^{2}\left({\pi_{8}^{\alpha}+\pi_{8}^{\alpha 3} \over \sqrt{2}}\right)=
{\Lambda^{2} \over 2 \pi^{2} v^{2}}\left( m_t^2+{9\over 4}v^2 g_s^2\right)
=(974~GeV)^2 \nn\\
& &M^{2}\left({\pi_{8}^{\alpha}-\pi_{8}^{\alpha 3} \over \sqrt{2}}\right)=
{\Lambda^{2} \over 2 \pi^{2} v^{2}}\left( m_b^2+{9\over 4}v^2 g_s^2\right)
=(930~GeV)^2 \nn\\
& &M^{2}\left( {P^{0i}_{3}+P_{3}^{3i}\over \sqrt{2}}\right)
={\Lambda^{2}
\over  2 \pi^{2} v^{2}}(4 m_t^2+v^2 g_s^2+{1\over 3}v^2 g'^2)
=(784~GeV)^2 \nn\\
& &M^{2}\left( {P^{0i}_{3}-P_{3}^{3i}\over \sqrt{2}}\right)=
{\Lambda^{2}
\over  2 \pi^{2} v^{2}}( 4 m_b^2+v^2 g_s^2+{1\over 3}v^2 g'^2)
=(560~GeV)^2 \nn\\
& &M^{2}\left( P_{3}^{-i}\right)=
{\Lambda^{2}
\over  2 \pi^{2} v^{2}}(3 m_t^2+m_b^2+v^2 g_s^2-{1\over 6}v^2 g'^2)
=(726~GeV)^2 \nn\\
& &M^{2}\left( P_{3}^{+i}\right)=
{\Lambda^{2}
\over  2 \pi^{2} v^{2}}(m_t^2+3 m_b^2+v^2 g_s^2+{5\over 6}v^2 g'^2)
=(632~GeV)^2
\eea
where $g_s$, $g$, $g'$ are the $SU(3)\otimes SU(2)_L\otimes U(1)_Y$ gauge
couplings, and $v\simeq 246 GeV$.

For the calculation we need the trilinear coupling of the $V$ to the PGB's
as derived in ref. \cite{bessu8}
\bea
{\cal L}^{(3)} & = &i g_{V\pi\pi}\Big\{V^3_\mu\Big
({\tilde\pi}^-\dmus{\tilde\pi}^+
+\pi_8^{\alpha -}\dmus\pi_8^{\alpha +}
-{\bar P}_3^{-i}\dmus P_3^{-i}+{\bar P}_3^{+i}\dmus P_3^{+i})\nn\\
& &+\frac {2}{\sqrt 3}V_{D\mu}
 ({\bar P}_3^{0i}\dmus  P_3^{0i}+{\bar P}_3^{3i}\dmus { P}_3^{3i}+
 {\bar P}_3^{-i}\dmus P_3^{-i}+{\bar P}_3^{+i}\dmus P_3^{+i})\nn\\
& &+V_\mu^-(\dmus{\tilde\pi}^3{\tilde\pi}^+
-{\tilde\pi}^3\dmus{\tilde\pi}^+
+\dmus \pi_8^{\alpha 3}\pi_8^{\alpha +}-
\pi_8^{\alpha 3}\dmus\pi_8^{\alpha +}
+\dmus P_3^{3i}{\bar P}_3^{-i}-{\bar P}_3^{3i}\dmus P_3^{+i})
\Big\}\nn\\
& &+h.c.
\eea
where
\be
g_{V\pi\pi}=\frac{\gs}{4}\frac{x^2}{r_V}\left(1-z^2\right)
\ee
with
\be
r_V=\frac {M^2_W}{M_V^2}~~x=\frac {g}{\gs}
\ee
The coupling constant $g''$ is the gauge coupling of the $V$ resonance
and $z$ is a combination of
free parameters appearing in front of the BESS lagrangian (see eq.(2.33) of
ref.\cite{bessu8}). The case $z=0$ corresponds to decoupling of the
axial-vector resonances.

The mixing of the new vector bosons $V^a$ can be directly read
in eq.(2.27) of ref.\cite{bessu8}. The elementary cross section is given by
\be
\frac {d\sigma}{d t}=\frac {1} {12} \frac {\vert M\vert^2}{16 \pi s^2}
\ee
For the process (2.1) we have
\bea
{\vert M\vert^2}&=&8 (ut -M_P^4)g_{V\pi\pi}^2
\Big\{(v_Z^2+a_Z^2)\frac{ T_{VZ}^2}{(s-M_Z^2)^2+M_Z^2\Gamma_Z^2)}\nn\\
&&+v_\gamma^2 \frac{ T_{V\gamma}^2}{s^2}\nn\\
&&+2 v_\gamma v_Z
\frac{T_{VZ}T_{V\gamma}}{(s-M_Z^2)^2+M_Z^2\Gamma_Z^2}
\frac {(s-M_Z^2)}{s}\Big \}
\frac {1}{(s-M_V^2)^2+M_V^2\Gamma_V^2}\nn\\
\eea
where
\be T_{VZ}=\frac {v^2}{4} g \gs \frac {x^2}{r_V}
\frac {\cos 2\theta}{\cos\theta}
{}~~~T_{V\gamma}=\frac {v^2}{2} g \gs \frac {x^2}{r_V}
\sin\theta
\ee
and $M_P\equiv M_{P^\pm}$.

The couplings of the gauge bosons $Z$ and $\gamma$ to the fermions are given
by
\be
v_\gamma=e Q~~~v_Z=\frac{e}{2\sin 2 \theta}(T_3-4Q \sin^2\theta)
{}~~~~~a_Z=\frac{e}{2\sin 2 \theta}T_3
\ee
with $T_3=\pm 1$.

For the process (2.2) we have
\bea
{\vert M\vert^2}&=&8(ut -{M_{P^\pm}}^2{M_{P^0}}^2)g_{V\pi\pi}^2
(v_W^2+a_W^2)\nn\\
&&\frac{ T_{VW}^2}{(s-M_W^2)^2+M_W^2\Gamma_W^2)}
\frac {1}{(s-M_V^2)^2+M_V^2\Gamma_V^2}
\eea
where
\be
T_{VW}=\frac {v^2}{4} g \gs \frac {x^2}{r_V}
\ee
with
\be
v_W=a_W=\frac{e}{2\sqrt{2}\sin \theta}
\ee
The resonant process is dominant, as it results for instance by
comparing the non resonant cross section
\be
\sigma(e^+e^-\rightarrow P^+P^-)=\frac {1}{4}(1-
\frac {4M_{P}^2}{s})^{3/2}
\sigma(e^+e^-\rightarrow \mu^+\mu^-)
\ee
with the results of our calculations (remember that, at $1~TeV$,
$\sigma(e^+e^-\rightarrow \mu^+\mu^-)=87~fb$).

\resection{PGB's at future linear $e^+e^-$ colliders}

Projects of $e^+e^-$ linear colliders are being studied at present at different
laboratories (among them SLAC, KEK, Novosibirsk and Serpukhov, DESY, Darmstadt,
CERN, with several other groups working in various universities). For general
reviews of such studies and of the physics potentialities we refer to
\cite{ee500} and \cite{saari}.

We have already discussed the usefulness of very energetic $e^+e^-$
colliders in exploring an alternative
scheme of electroweak symmetry breaking based on a strong interacting sector
with new vector resonances \cite{bessee}.
The sensitivity of $e^+e^-$ linear colliders, for different proposed options of
energies and luminosities, to the BESS model parameters, was there
quantitatively examined, particularly in connection to the vector bosons
$V$ that appear in the model.

The $V$'s can be produced as real resonances if their mass is below
the collider energy. In a high energy collider one expects to see dominant
peaks below the maximum c.m. energy corresponding to such resonances. Due to
beamstrahlung and synchrotron radiation it may not become necessary to tune the
beam energies in order to see such peaks.

In ref. \cite{bessu8} we were
essentially interested in the neutral vector resonances of the strongly
interacting sector as described by BESS. The description of such resonances in
BESS is rather general, and, after convenient choice of parameters may also
apply to a phenomenological description of the standard techni-$\rho$.
Studies of the techni-$\rho$ production in $e^+e^-$ colliders have been
carried out by Peskin
\cite{Peskin}, Iddir et al. \cite{Iddir}, through study of strong final state
interaction, and have also been discussed by Barklow \cite{Barklow} and Hikasa
\cite{Hikasa}, via various methods.

A prediction of such models for the electroweak breaking, in the general case
of a large initial global symmetry group, is the existence of pseudo-Goldstone
bosons. Our calculations here are within extended BESS.
Future $e^+e^-$ colliders are one of the best opportunities to study the
production of pairs of charged pseudo-Goldstone bosons. The methods used for
their detection turn out to be very similar to those used in the case of
charged Higgs searches \cite{Gunion}, but with the advantage that PGB can
here be resonantly produced: $e^+e^- \to V \to P^+P^-$. From now on we shall
use the following notations:
$P^\pm\equiv {\tilde \pi}^{\pm}$,
and  $P^0\equiv
{\dd{\tilde\pi}^{3}-\pi_{D}\over \dd\sqrt{2}}$
for the lightest neutral PGB's.

The peak cross section is
\be
\sigma(M_V^2)=12 \pi \frac {\Gamma_V^e \Gamma^P_V}
{M_V^2 (\Gamma_V^{TOT})^2}
\ee
with
\be
\Gamma_V^e= \frac{4}{3}\alpha_{em} M_V (v_e^2+a_e^2)
\ee
where we have used the couplings of the $V$ to the fermions \cite{BESS}
\be
v_f=\frac{1}{2\sin 2\theta}(C T_3+4 D Q)~~
a_f=\frac{1}{2\sin 2\theta}C T_3
\ee
with
\be
C=-\frac {\cos 2 \theta}{\cos\theta}\frac{g}{\gs}~~~
D=-\frac {\sin^2\theta}{\cos\theta}\frac{g}{\gs}
\ee

The decay width of $V^3$ in $P^+P^-$ is given by
\be
\Gamma_V^P=\frac {1}{48\pi}g^2_{V\pi\pi}M_V(1-4 \frac{M_P^2}{M_V^2})^{3/2}
\ee
In computing $\Gamma_V^{TOT}$ we have neglected the fermion contribution.

We present in the following table  the results for the total cross section for
$M_V=1~TeV$ and $g/\gs=0.05$ (this value is compatible with the present
limitations coming from LEP1 \cite{Grazz}). The assumed spectrum for the
PGB's is that of eq.(2.4).
\begin{center}
\begin{tabular}{|c|c|c|c|c|}
\hline
z&$\sigma(pb)$&$\Gamma_V^W$&$\Gamma^P_V(GeV)$&$\Gamma_V^T(GeV)$\\
\hline
0   & 1.6 & 10.3 & 2.7 & 13  \\
0.5 & 2.8 & 5.8  & 1.5 & 7.3 \\
0.7 & 6.1 & 2.7  & 0.7 & 3.4 \\
\hline
\end{tabular}
\end{center}
Notice that the peak cross sections do not depend on $g/\gs$,
and they increase with $z$ as $1/(1-z^2)^2$. The large cross sections
that we have found are due to the relatively small $V$ mass assumed,
and to the large branching ratio of $V\rightarrow
P^+P^-$. In the following table we exhibit the sensitivity to
the $V$ mass (assuming the worst case, $z=0$). The cross section decreases
very rapidly when $V$ becomes heavier.
\begin{center}
\begin{tabular}{|c|c|c|c|c|c|c|c|}
\hline
$\sqrt {s}=M_V~(TeV)$ & $\sigma(pb)$ & $\Gamma_V^W$ & $\Gamma^P_V(GeV)$ &
$\Gamma_V^T(GeV)$ & $\sigma_{WW}(pb)$ & $\sigma_{ZZ}(pb)$ &
$\sigma_{t\bar t}(pb)$\\
\hline
1.0 & 1.6  & 10.3  & 2.7   & 13    & 2.7+6.1 & 0.21 & 0.20 \\
1.2 & 0.69 & 25.9  & 11.8  & 37.7  & 2.1+1.5 & 0.16 & 0.14 \\
1.5 & 0.11 & 79.7  & 50.8  & 172.4 & 1.4+0.2 & 0.12 & 0.10 \\
1.7 & 0.03 & 149.6 & 106.8 & 456.5 & 1.2+0.04 & 0.10 & 0.08 \\
\hline
\end{tabular}
\end{center}
The background processes in the above table are $e^+e^-\rightarrow t\bar t$,
$e^+e^-\rightarrow W^+W^-$, $e^+e^-\rightarrow ZZ$. With respect
to the standard model we have for $\sigma_{WW}$ an enhancement (given in the
sixth column of the table as the second number) due to the
production of a couple of $W$'s from the $V$ resonance, while for
$\sigma_{ZZ}$ the numerical value is that of the standard model
as the coupling $V^3ZZ$ is zero. For $\sigma_{t\bar t}$ the numerical value
is again very close to that of the standard model, as, excluding a direct
coupling of $V$ to fermions, the extra contributions with respect to the
standard model are suppressed by the small mixing factor $g/g''$.

Concerning the decay of the PGB's we have
\be
\Gamma(P^+\rightarrow t\bar {b})=\frac {1}{8 \pi}
\frac {m_t^2}{v^2} M_P \sqrt {1- \frac{m_t^2}{M_P^2}}
\ee

Therefore we have to consider the following final state
\bea
P^+P^-\rightarrow t\bar{ b}\bar {t}b
\rightarrow Wb\bar{ b}W\bar {b}b
\rightarrow jjb\bar{ b}l\nu\bar {b}b
\eea

These backgrounds have already been considered
for the charged Higgs boson production at future $e^+e^-$
colliders \cite{Sopczak}. We are a priori in a more favorable situation
since the PGB we have studied are resonantly produced. A detailed analysis
of these processes requires full knowledge of the experimental set-up and
is beyond the scope of this work. Nevertheless in our case the signal
cross-section is, even in the case z=0, of the same order of magnitude and in
some cases even larger than the background, thus favoring in principle signal
detection.

The background process
$e^+e^-\rightarrow W^+W^-$ can be easily reduced below 1\% of its initial value
by requiring the tagging of one $b$ in the final state. This can be easily
understood, since the branching ratio of $W \to b\; +\; u$-type quark
is very low due to either a small Kobayashi-Maskawa matrix element or
phase-space suppression.

In principle also invariant mass and $p_T$ cuts may be useful,
but in order to reconstruct a PGB mass or $p_T$ it is necessary to face a
many-jet problem, i.e. jet-combinatorics and isolation. Moreover not only the
signal, but also $WW$ production, is a high $p_T$ process, thus reducing
considerably the efficiency of a $p_T$ cut.

Similar considerations apply to
the $e^+e^-\rightarrow ZZ$ background. In this case the tagging of one $b$ is
less efficient in reducing the background, but the cross-section is
already more than one order of magnitude smaller than the preceeding one.

The background $e^+e^-\rightarrow t\bar t$ is more difficult to weed out.
The study of charged Higgs boson discovery potential has shown that a
microvertex detector is crucial for establishing the signal over the
$e^+e^-\rightarrow t\bar t$ background \cite{Sopczak}.

We will assume $B=0.20$ for the
product of the branching ratios $W\rightarrow hadrons$ and
$W\rightarrow leptons$, and a $b$-tagging efficiency $\eps_b=0.5$.
Assuming an integrated luminosity of $80~fb^{-1}$, after multiplication of
the number of events of the Table by the branching ratios and  $b$-tagging
efficiencies, a still large number of events  is left when $M_V=1~TeV$. When
$M_V$ is higher than $1.2~TeV$ and $z=0$, the signal becomes smaller than
background and deserves careful study of background rejection.

To conclude this section we mention the suggestion, originally due to Ginzburg
et al. \cite{Ginzburg}, of the possibility of
obtaining an energetic photon beam by colliding
electron bunches with a laser beam in the visible spectrum. This
technique should allow keeping a high luminosity for the photon beam from the
back scattered laser, and it should allow for a photonic spectrum mostly
concentrated at the highest energies, not much lower than the electron energy.
Such a behaviour is quite different from that of the expected beamstrahlung
photons concentrated at the lower energies, and from that of the bremstrahlung
photons. This technical possibility would thus allow for energetic
photon-photon and electron-photon collisions.
For the purpose of the present paper, where we
are interested in a possible strong electroweak sector, $\gamma \gamma$
collisions would appear of interest if resonant behaviours are present in
states of zero angular momentum which can couple to two real photons. These
states may be pseudo-Goldstone states, when one considers the effective
interaction due to the Adler-Bell-Jackiw anomaly. Also interesting will be
the production of pairs of such pseudos from $\gamma \gamma$ (both real or
virtual). If energetically reachable, however, the
resonant behaviours we have discussed will be more prominent signals.

\resection{PGB's at LHC}

Much work has been devoted in recent years to study of physical implications of
a possible strong electroweak sector in view of future hadronic colliders
\cite{strongint}.

We have considered the PGB production at LHC via the charged $V$
resonance. Also in this case the methods used for PGB detection are similar
to those used in the case of charged Higgs \cite{Gunion}, but with the
advantage that the PGB are produced from a charged $V$ resonance. We have
evaluated the quark-antiquark annihilation (Drell-Yan type)
contribution to the differential cross section both in terms of the
invariant mass of the pair of PGB's and in terms of the transverse momentum
$p_T$ of the PGB for the process
\be
pp\rightarrow W^\pm\rightarrow V^\pm\rightarrow P^\pm P^0+X
\ee
Its expression follows directly from the partonic
cross section given in eq.(2.2). The partial width of
the decay $V^\pm\rightarrow P^\pm P^0$
is
\be
\Gamma_{V\pm}=\frac {1}{96\pi}g^2_{V\pi\pi}M_V
\Big [(1- \frac{(M_{P^\pm}+M_{P^0})^2}{M_V^2})
(1- \frac{(M_{P^\pm}-M_{P^0})^2}{M_V^2})\Big ]^{1/2}
[1-2\frac{(M_{P^\pm}^2+M_{P^0}^2)}{M_V^2}]
\ee
We have then added the fusion contribution coming from the process
\be
pp\rightarrow W^\pm Z\rightarrow V^\pm\rightarrow P^\pm P^0
\ee
The fusion amplitudes have been computed using the equivalence
theorem \cite{equiv}.

To compute $W^+Z\rightarrow P^+P^0$ we need
${\cal L}_{\pi\pi{\tilde \pi}{\tilde \pi}}$ \cite{bessu8}:
\bea
{\cal L}_{\pi\pi{\tilde \pi}{\tilde \pi}}&=&
-\frac {1}{6} \frac {1}{v^2}(1-\frac {3}{4}\alpha)
[ -{\dmu{\tilde\pi}}^+\dmus{{\tilde\pi}}^+(\pi^-)^2
+{\dmu{\tilde\pi}}^+\dmus{{\tilde\pi}}^-\pi^+\pi^-\nn\\
& &+{\dmu{\tilde\pi}}^3\dmus{{\tilde\pi}}^3\pi^+\pi^-
-2{\dmu{\tilde\pi}}^3\dmus{{\tilde\pi}}^+\pi^3\pi^-
+{\dmu{\tilde\pi}}^+\dmus{{\tilde\pi}}^-(\pi^3)^2\nn\\
& &-4{\dmu{\tilde\pi}}^+{{\tilde\pi}}^-\dmus\pi^+\pi^-
+4{\dmu{\tilde\pi}}^-{{\tilde\pi}}^3\dmus\pi^+\pi^3
+4{\dmu{\tilde\pi}}^3{{\tilde\pi}}^-\dmus\pi^3\pi^+\nn\\
& &+2{\dmu{\tilde\pi}}^-{{\tilde\pi}}^-\dmus\pi^+\pi^+
+2{\dmu{\tilde\pi}}^+{{\tilde\pi}}^-\dmus\pi^-\pi^+
-({{\tilde\pi}}^-)^2\dmu\pi^+\dmus\pi^+\nn\\
& &
+{{\tilde\pi}}^-{{\tilde\pi}}^+\dmu\pi^+\dmus\pi^-
+{{\tilde\pi}}^-{{\tilde\pi}}^+\dmu\pi^3\dmus\pi^3
-2{\dmu{\tilde\pi}}^3{{\tilde\pi}}^3\dmus\pi^+\pi^-\nn\\
& &-2{\dmu{\tilde\pi}}^+{{\tilde\pi}}^3\dmus\pi^3\pi^-
-2{{\tilde\pi}}^-{{\tilde\pi}}^3\dmu\pi^+\dmus\pi^3
+(\tilde\pi_3)^2\dmu\pi^+\dmus\pi^-\nn\\
& &-2{\dmu{\tilde\pi}}^+{{\tilde\pi}}^-\dmus\pi^3\pi^3
-2{\dmu{\tilde\pi}}^3{{\tilde\pi}}^-\dmus\pi^+\pi^3
+h.c.]
\eea
The fusion amplitude for the process $W^+Z\rightarrow P^+P^0$ is given by
\be
A(W^+Z\rightarrow P^+P^0)=\frac{1}{\sqrt{2}}
i \lq (1-\frac {3}{4}\alpha)\frac{t}{v^2}+
\frac {\alpha}{4}\frac{M_V^2}{v^2}
(\frac {u-t}{s-M_V^2}+\frac{s-t}{u-M_V^2})\rq
\ee

We have also evaluated the amplitude $W^+Z \to W^+Z$ through the equivalence
theorem, using the trilinear coupling, given in eq.(2.5), the quadrilinear
coupling being given by \cite{bessu8}
\bea
{\cal L}_{\pi\pi\pi\pi}&=&-\frac {1}{6} \frac {1}{v^2}
(1-\frac {3}{4}\alpha)
[ -{\dmu\pi}^+\dmus{\pi}^+(\pi^-)^2\nn\\
& &+{\dmu\pi}^+\dmus{\pi}^-\pi^+\pi^-
+{\dmu\pi}^3\dmus{\pi}^3\pi^+\pi^-\nn\\
& &-2{\dmu\pi}^3\dmus{\pi}^+\pi^3\pi^-
+{\dmu\pi}^+\dmus{\pi}^-(\pi^3)^2+h.c.]
\eea
where
\be
\alpha=\frac {x^2}{r_V} (1-z^2)^2
\ee

The result, to be compared
with our previous work based on $SU(2)\otimes SU(2)$,
is
\be
A(W^+Z\rightarrow W^+Z)=i
\lq (1-\frac {3}{4}\alpha)\frac{t}{v^2}+
\frac {\alpha}{4}\frac{M_V^2}{v^2}
(\frac {u-t}{s-M_V^2}+\frac{s-t}{u-M_V^2})\rq
\ee

In Fig. 1 (Fig. 2) we plot the invariant mass ($p_T$) distribution for the
set of parameters $M_V=1000~GeV$, $\gs=13$ and $z=0$; we have added the events
of the $P^+P^0$ channel with the events of the $P^-P^0$ one. We have plotted
in Fig. 3 and 4 the same distributions with $M_V=1200~GeV$, $\gs=6.9$ and
$z=0.5$.

The invariant mass distributions show a peak around
the mass of the $V$, and the $p_T$ distribution
is characterized by a jacobian peak, the broadness
being directly related to the $V$ width.
The fusion subprocess gives a small contribution (less than roughly
$10 \%$).

We can perform the same analysis also for the neutral channel:
\be
pp\rightarrow Z,\gamma\rightarrow V^3\rightarrow P^+ P^-+X
\ee

In this case, the fusion amplitudes are given by

\be
A(W^+W^-\rightarrow P^+P^-)=i
\lq (1-\frac {3}{4}\alpha)\frac{u}{v^2}+
\frac {\alpha}{4}\frac{M_V^2}{v^2}
(\frac {t-u}{s-M_V^2}+\frac{s-u}{t-M_V^2})\rq
\ee
and

\be
A(ZZ\rightarrow P^+P^-)=i
\lq (1-\frac {3}{4}\alpha)\frac{-s}{v^2}+
\frac {\alpha}{4}\frac{M_V^2}{v^2}
(\frac {s-u}{t-M_V^2}+\frac{s-t}{u-M_V^2})\rq
\ee
Notice that in the $t$ channel of the amplitude (4.10) and in the $t$ and $u$
channels of (4.11) we have to consider the exchange of the triplet $\tilde V$,
which however in BESS is practically degenerate in mass with the triplet of
the $V$. As shown in Fig. 5 and 6 for $M_V=1000~GeV$, $\gs=13$ and $z=0$, the
total number of events is depressed by roughly one order of magnitude compared
to the other case, since the Drell-Yan type subprocess is very small.
Therefore the fusion subprocess becomes important (roughly $30 \%$).
Concerning the final state, the decay widths $P^+$ are given in eq.(3.6),
while the partial decay width of $P^0$ in $\bar {b}b$ is given by
\be
\Gamma(P^0\rightarrow b\bar {b})=\frac {1}{8 \pi}
\frac {m_b^2}{v^2} M_P \sqrt {1- \frac{4~m_b^2}{M_P^2}}
\ee
to be compared with the $P^0$ decay to two gluons \cite{Lubicz}
\be
\Gamma(P^0\rightarrow gg)=\frac {\alpha_s^2}{8 \pi^3}
\frac {N_{TC}^2}{6v^2} M_P^3
\ee
where $N_{TC}$ is the number of technicolor. For masses of $P^0$ in the
considered range this decay is less important. The $P^0$
decay in two photons is depressed by a factor $(\frac {\alpha}
{\alpha_s})^2$.

Therefore the expected signals for $P^\pm P^0$ are
$t\bar {b}b\bar {b}$ or $t\bar {b}gg$, and
 $t\bar {t}b\bar {b}$ for $P^+P^-$. These final states have
to be studied and compared with the background.

For instance, the $t\bar {t}b\bar {b}$ background comes from
the following processes
\bea
&&gg\rightarrow  t\bar {t}b\bar {b}\nn\\
&&gg\rightarrow  t\bar{t}Z\rightarrow t\bar {t}b\bar {b}\nn\\
&&gg\rightarrow  t\bar{t}+jets
\eea
where the last jets are misidentified as $b$ jets. These background have been
studied \cite{Gunion} for charged Higgs boson discovery from $tb$ decays at
LHC using SDC detector. In order to disentangle signal from background,
reasonably efficient and pure $b$ tagging is mandatory. We recall that the
energy of LHC is now taken to be at 14 TeV. The following procedures have been
applied in ref. \cite{Gunion}. A lepton with high momentum
($p_T>20~GeV$), coming from the $t$ decays (via $W$) is used as trigger
and the missing momentum $p_T>50~GeV$ required.
The second $W$ coming from the second $t$ is assumed to decay hadronically.
The invariant mass of each pair of jets (not containing tagged $b$
quarks) is required to satisfy
\be
M_W-\frac {\Delta M_W}{2}<M_{jj}<
M_W+\frac {\Delta M_W}{2}
\ee
Then for each pair satisfying the previous criterion
one computes a three jet invariant mass $M_{bjj}$,
by combining $M_{jj}$ with a tagged $b$ jet, and requires
that
\be
m_t-\frac {\Delta m_t}{2}<M_{bjj}<
m_t+\frac {\Delta m_t}{2}
\ee
Using this cuts one can reduce the background.
Finally one can compute $M_{bbjj}$ and make a plot of
the $M_{bbjj}$ invariant mass distribution.
Clear signal peaks appear except when $M_{H^+}\sim m_t$.

Our case deserves careful study along the previous procedure,
since the expected number of signal events is not large.
A similar analysis has also to be performed for the
$tbbb$ and $tbgg$ final states. We plan to do this
in the next future in collaboration with experimentalists involved
in LHC. Notice that it will be crucial to know if $b$
tagging can be performed with the planned huge luminosity
of $10^{34} cm^{-2} s^{-1}$.

Finally we have computed also the $WZ$ production which has
been shown to allow for identification of a charged $V$
at LHC up to $2~TeV$ masses within the $SU(2)\otimes SU(2)$ model.
Since new channels for $V^\pm$ decays are open, the signal
will be reduced. Three contributions coming from Drell-Yan type, fusion,
and SM background are summed up. In order to allow for a direct comparison
with previous minimal BESS \cite {LHC}
we have taken an LHC energy of $16~ TeV$. The rate of events decreases
by roughly $20 \%$ if we consider the presently planned energy of $14~ TeV$.

Fig. 7 and Fig. 8 (resp. Fig.  9 and Fig. 10)
give the prediction for invariant $WZ$
mass and $p_T$ of the $Z$ for $M_V=1000~GeV$, $\gs=13$
and $z=0$ for BESS $SU(2)$ (respectively for BESS $SU(8)$).
Although the signal is reduced by $20\%$ the identification
will be very easy. For higher masses, since the width increases
and the cross section for the production of PGB decreases,
the discovery potential is reduced down to $1.5~TeV$, as
shown in Fig. 11 and Fig. 12. In particular the shape of
the jacobian peak does not differ from the background, the
signal leading only to an excess of events.

Since only leptonic
decays of electroweak gauge bosons have been taken into account,
the situation may be less pessimistic, if identification
through hadronic $W$ decay and leptonic $Z$ decay, which has
been previously shown to be crucial for removing top
background, is possible. Within the ability
to perform $b$ tagging with good efficiency and purity,
one may hope to disentangle $t\bar t$ background from $W$ pair production
which is enhanced by $V^3$ resonance in our case.

\resection{Conclusion}

We have investigated the production of the lightest pseudo-Goldstone bosons
at future colliders through the resonances of the extended BESS model, which is
an effective lagrangian parametrization to describe the low energy
phenomenology of a general class of theories with dynamical symmetry
breaking. The existence of pseudo-Goldstone bosons is in fact a quite common
and interesting prediction of such theories.

Detection at LHC of a charged vector resonance through its decay into $WZ$
pairs is possible in the framework of the extended BESS model for a
significant domain of its parameter space. Production of pairs of
pseudo-Goldstone bosons $P^\pm P^0$ is also important, but discovery via
$tbbb$ or $ttgg$ decays needs a careful evaluation of backgrounds in the LHC
environment.

A more promising preview is instead obtained for production of charged
pseudo-Goldstones at the $V$ resonance in $e^+e^-$ collisions in the TeV range.
In fact the largest background, namely $WW$ production, can be easily reduced
to a very low level by requiring the tagging of one $b$ in the final state.
Other backgrounds, such as $ZZ$ and $t\bar t$ production, have smaller
cross-sections
as compared with signal cross-section, at least in a range of the parameter
space of the model. For increasing values of the $M_V$ mass and decreasing
values of the $z$ parameter (we have examined in detail the worst case $z=0$)
the signal cross-section becomes smaller than background and deserves a
detailed study of background rejection.

\newpage
\begin{center}
  \begin{Large}
  \begin{bf}
  Figure Captions
  \end{bf}
  \end{Large}
  \vspace{5mm}
\end{center}
\begin{description}
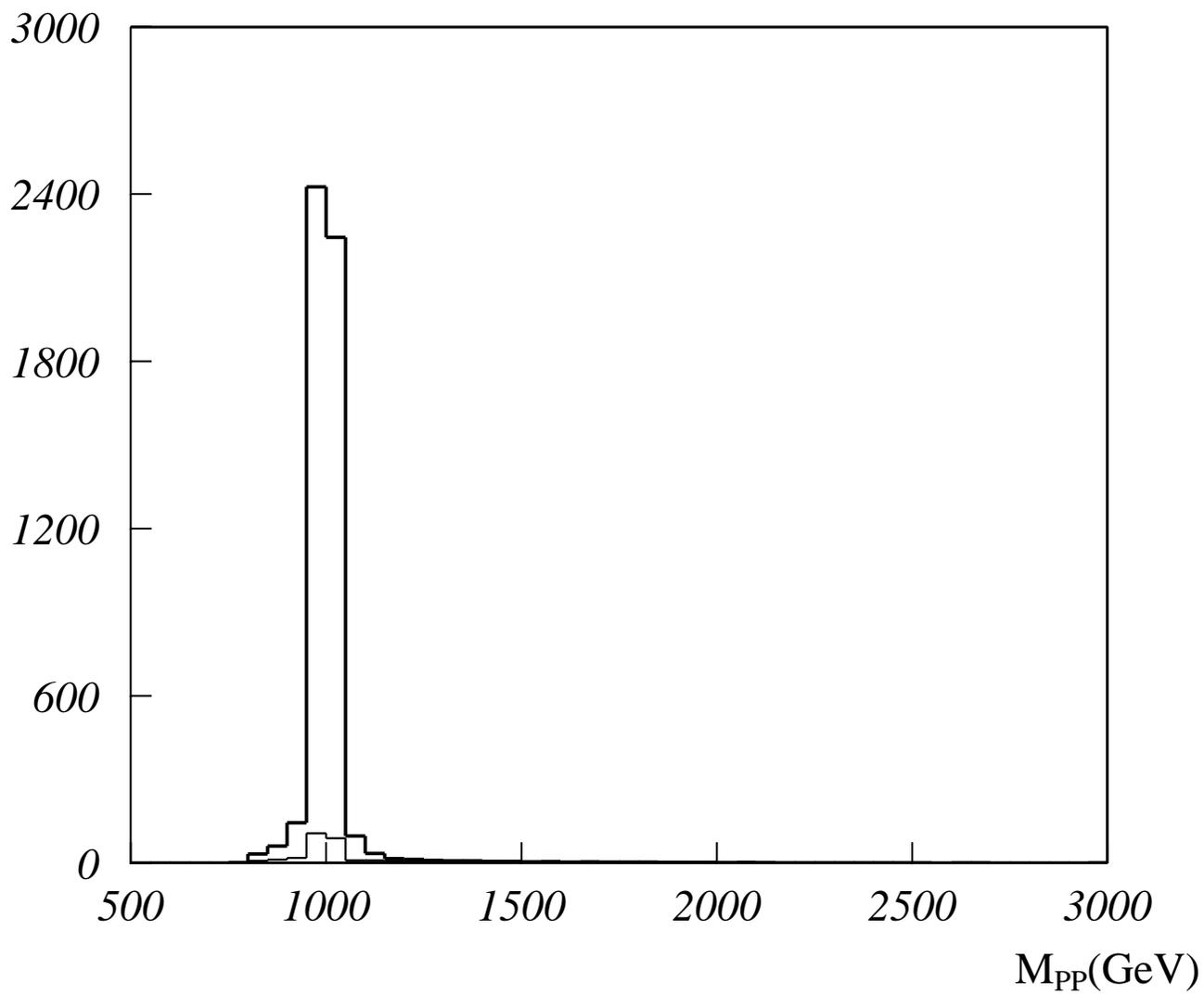
\item [Fig. 1] Invariant mass distribution of the $P^+P^0+P^-P^0$
produced per year at LHC for $M_V=1000~GeV$, $g''=13$ and $z=0$, with a
luminosity of $100~fb^{-1}$.
The applied cuts are: $|y_{P}|<2.5$, $(p_T)_{P^0}>300~GeV$. The lower (higher)
histogram refers to the fusion signal (fusion signal plus $q\bar q$
annihilation signal).

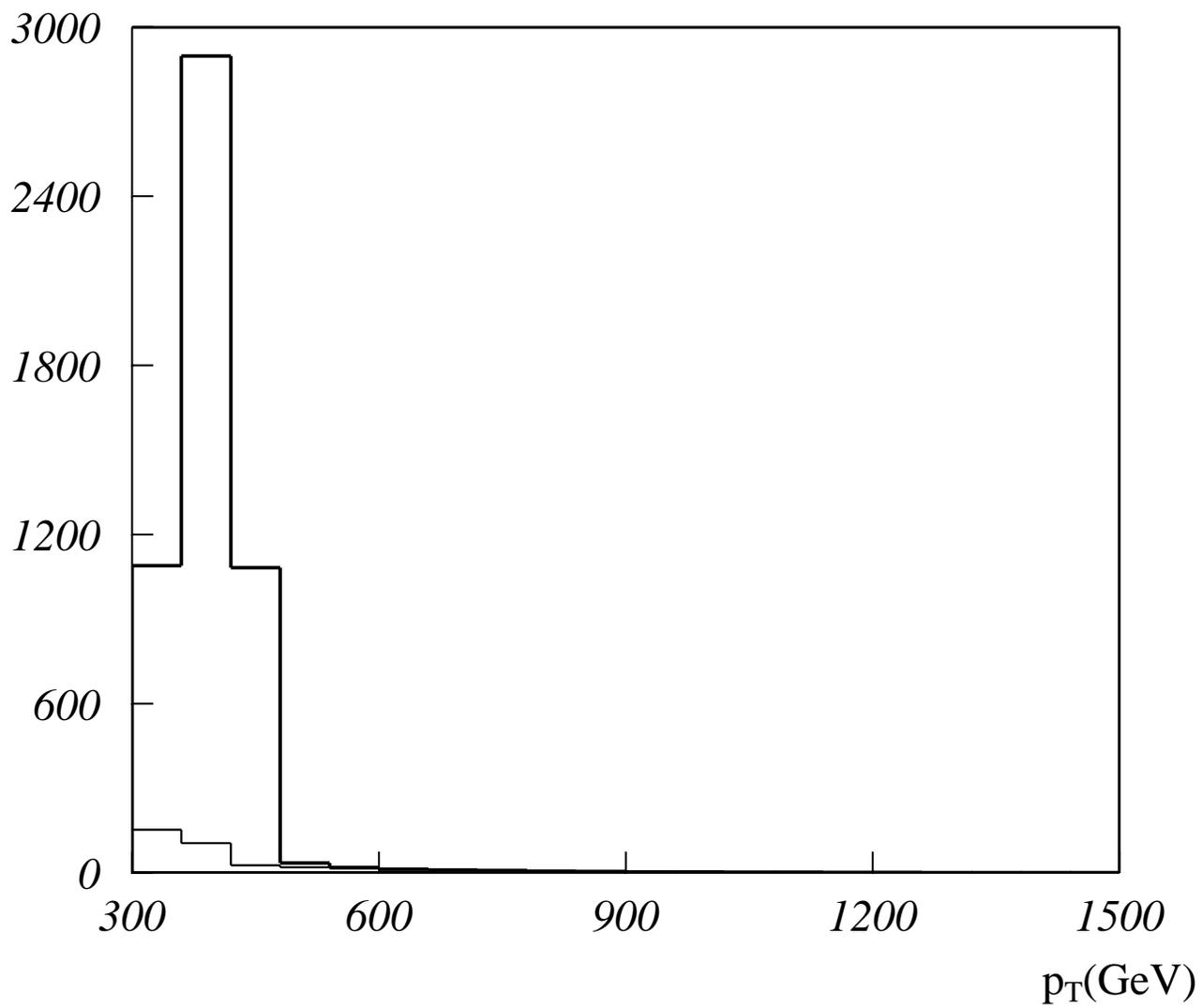
\item [Fig. 2] $(p_T)_{P^0}$  distribution of the $P^+P^0+P^-P^0$
               produced per year at LHC for $M_V=1000~GeV$,
               $g''=13$ and $z=0$, with a
               luminosity of $100~fb^{-1}$.
               The applied cuts are: $|y_{P}|<2.5$, $M_{PP}>500~GeV$.
               The lower  (higher) histogram
               refers to the  fusion signal
               (fusion signal plus $q\bar q$ annihilation signal).

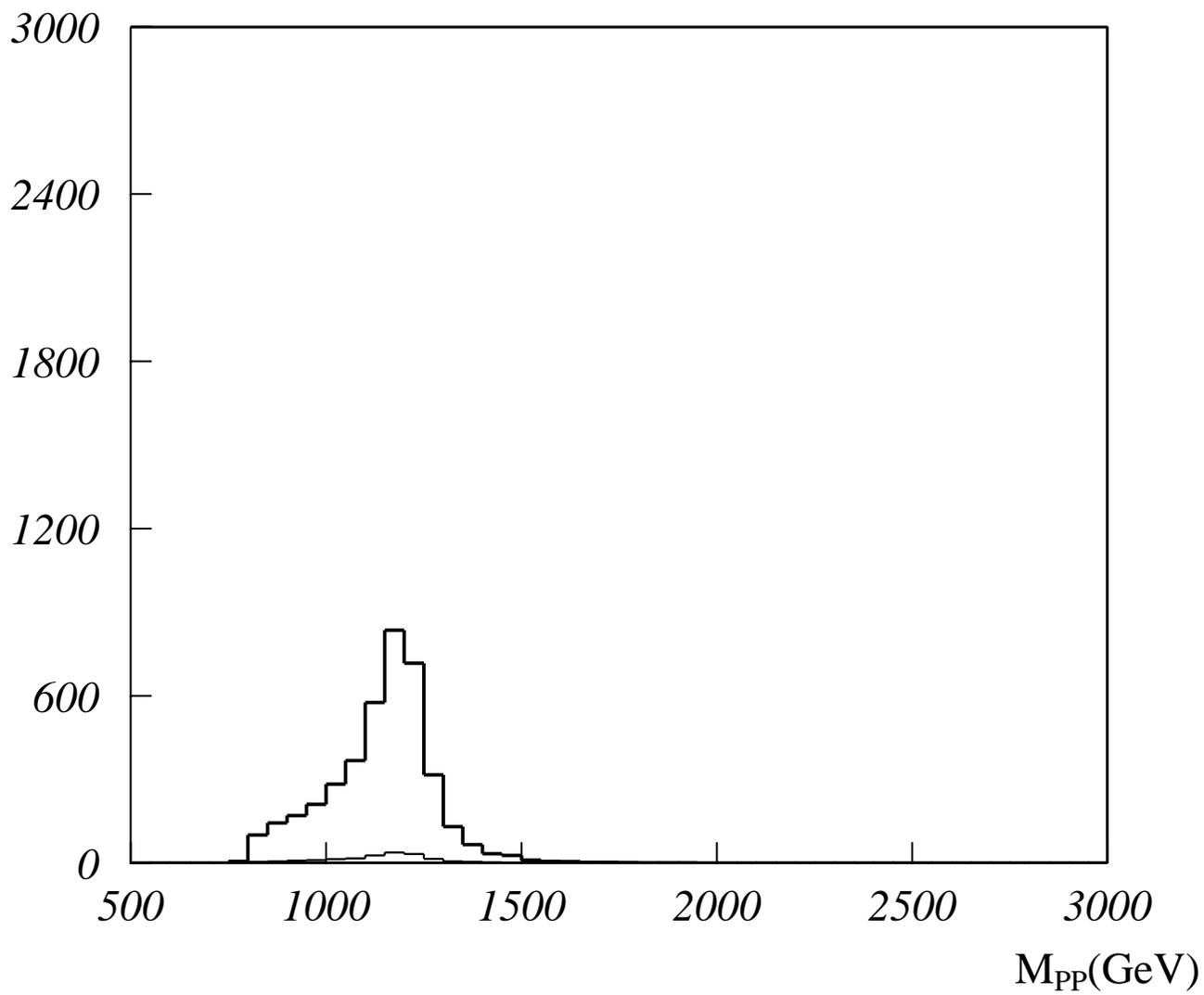
\item [Fig. 3] Invariant mass distribution of the $P^+P^0+P^-P^0$
               produced per year at LHC for $M_V=1200~GeV$,
               $g''=6.9$ and $z=0.5$, with a
               luminosity of $100~fb^{-1}$.
               The applied cuts are: $|y_{P}|<2.5$, $(p_T)_{P^0}>300~GeV$.
               The lower  (higher) histogram
               refers to the  fusion signal
               (fusion signal plus $q\bar q$ annihilation signal).

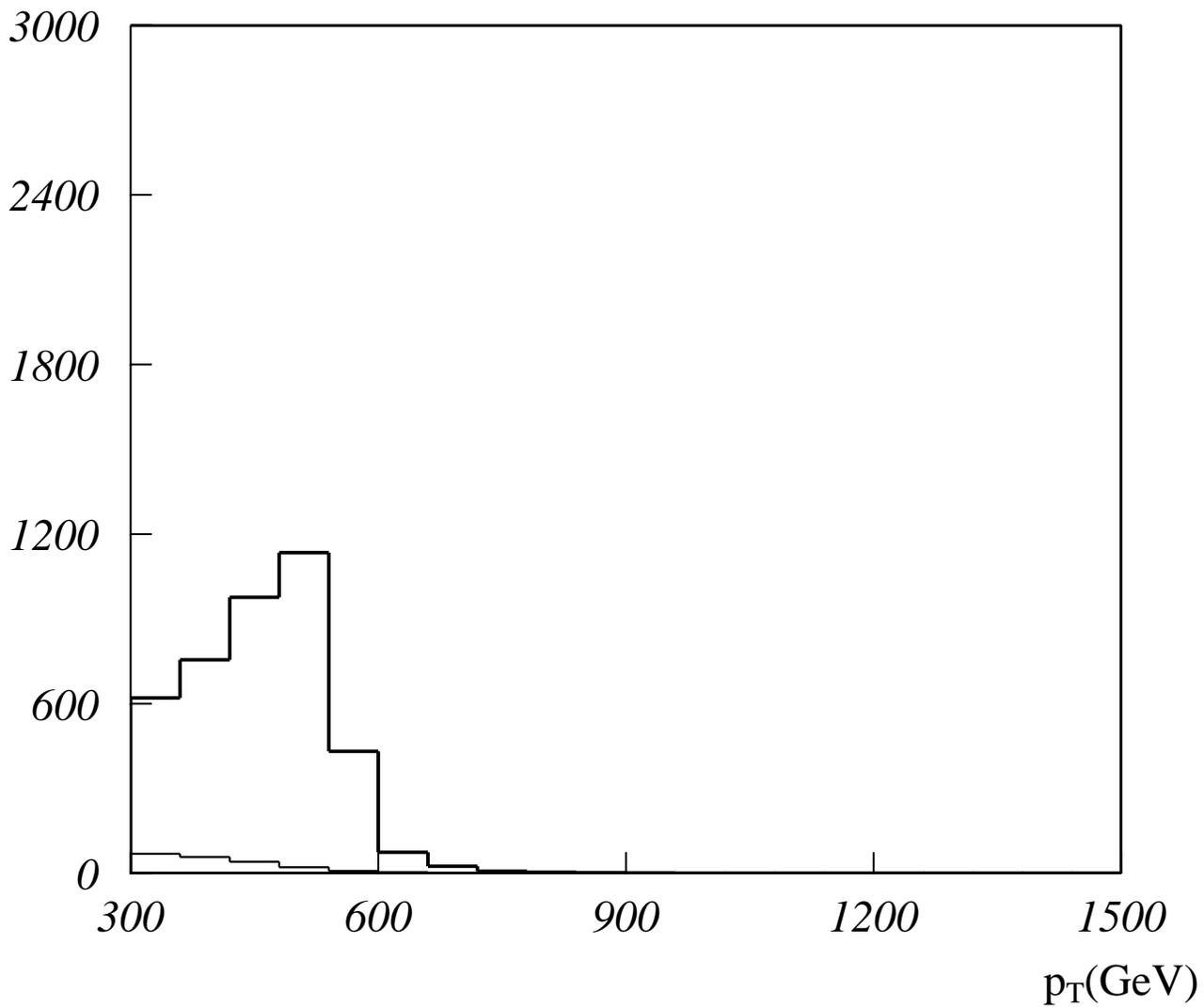
\item [Fig. 4] $(p_T)_{P^0}$  distribution of the $P^+P^0+P^-P^0$
               produced per year at LHC for $M_V=1200~GeV$,
               $g''=6.9$ and $z=0.5$, with a
               luminosity of $100~fb^{-1}$.
               The applied cuts are: $|y_{P}|<2.5$, $M_{PP}>500~GeV$.
               The lower  (higher) histogram
               refers to the  fusion signal
               (fusion signal plus $q\bar q$ annihilation signal).

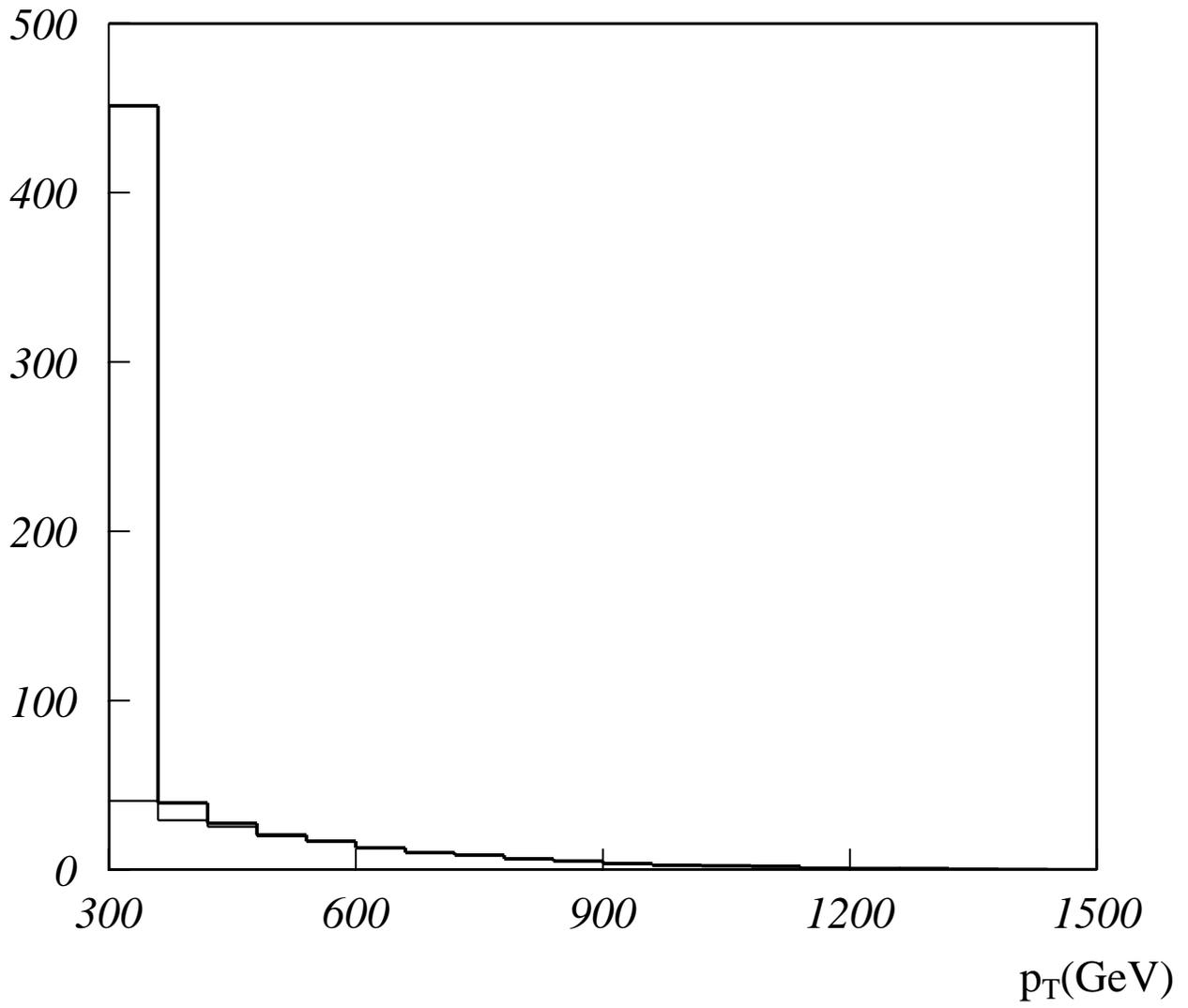
\item [Fig. 5] Invariant mass distribution of the $P^+P^-$
               produced per year at LHC for $M_V=1000~GeV$,
               $g''=13$ and $z=0$, with a
               luminosity of $100~fb^{-1}$.
               The applied cuts are: $|y_{P}|<2.5$, $(p_T)_{P^0}>300~GeV$.
               The lower  (higher) histogram
               refers to the  fusion signal
               (fusion signal plus $q\bar q$ annihilation signal).

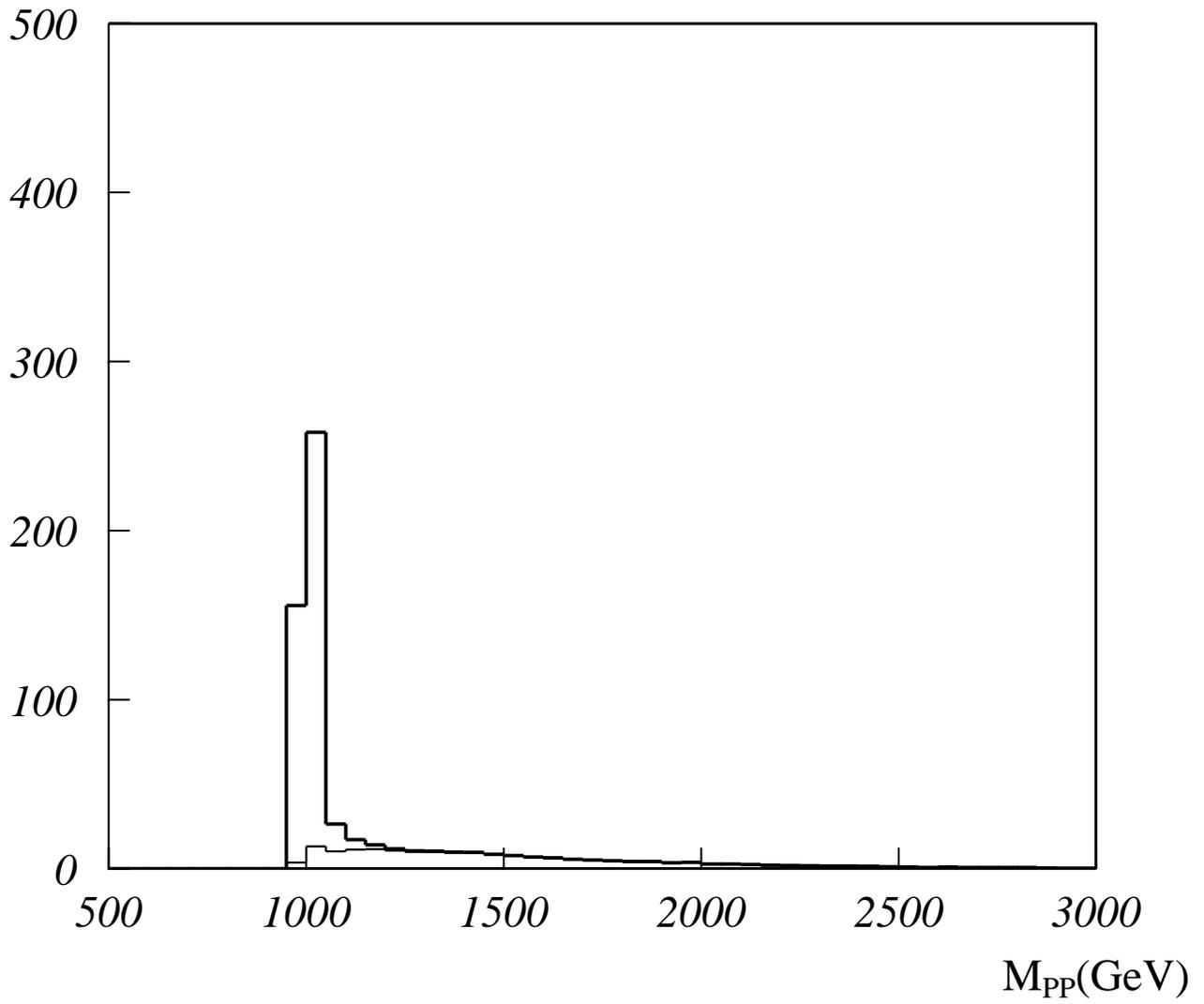
\item [Fig. 6] $(p_T)_{P^0}$  distribution of the $P^+P^-$
               produced per year at LHC for $M_V=1000~GeV$,
               $g''=13$ and $z=0$, with a
               luminosity of $100~fb^{-1}$.
               The applied cuts are: $|y_{P}|<2.5$, $M_{PP}>500~GeV$.
               The lower  (higher) histogram
               refers to the  fusion signal
               (fusion signal plus $q\bar q$ annihilation signal).

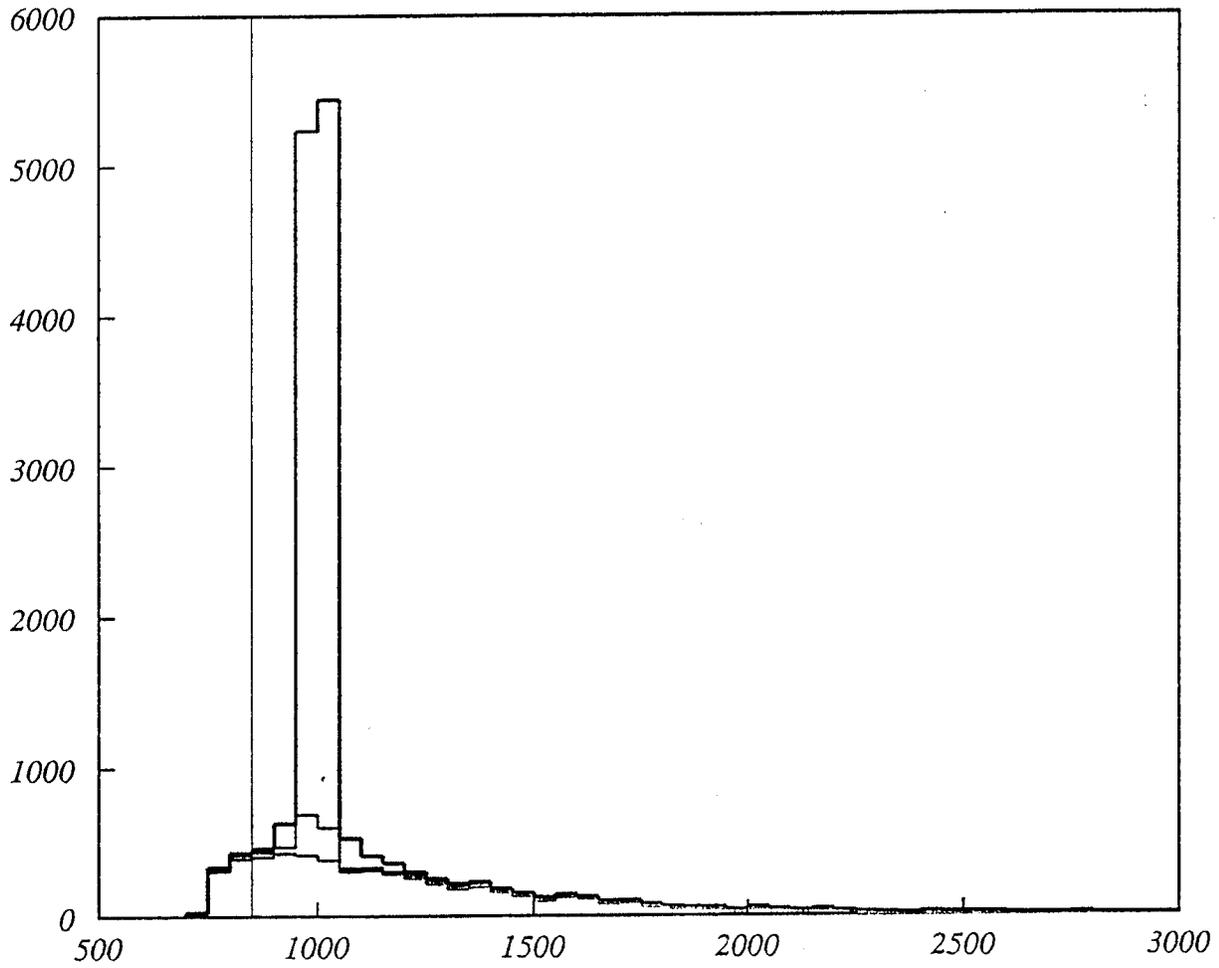
\item [Fig. 7] Invariant mass distribution of the $W^+Z+W^-Z$
               pairs produced per year
               at LHC $(\sqrt s =~16~TeV)$ for $M_V=1000~GeV$, $g''=13$
               and $z=0$ within BESS $SU(2)\otimes SU(2)$, with a
               luminosity of $100~fb^{-1}$.
               The applied cuts are: $|y_{W,Z}|<2.5$, $(p_T)_Z>360~GeV$
               and $M_{WZ}>850~GeV$.
               The lower, intermediate and higher histograms
               refer to the background, background plus fusion signal and
               background plus
               fusion signal plus $q\bar q$ annihilation signal, respectively.

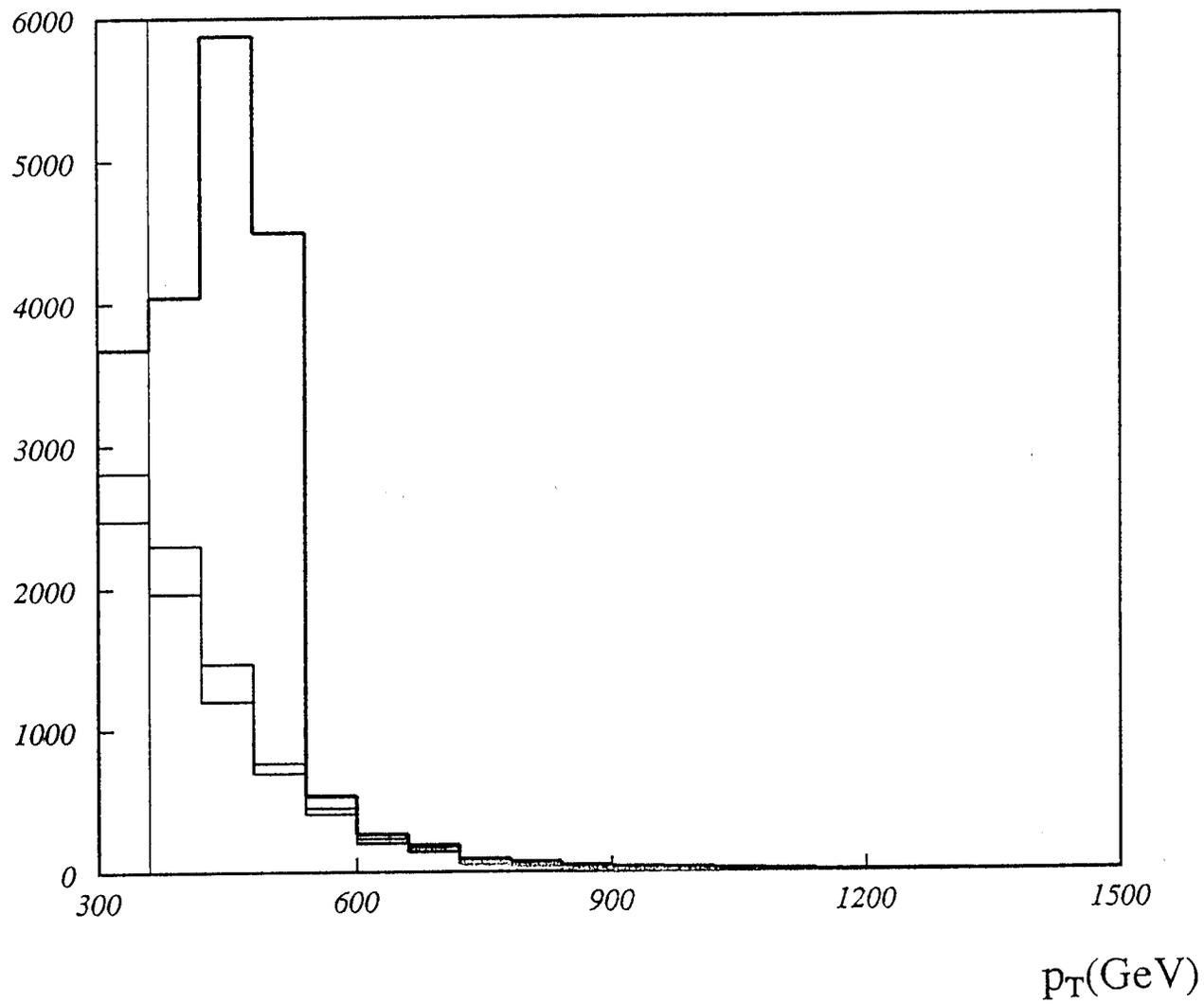
\item [Fig. 8] $(p_T)_Z$ distribution of the $W^+Z+W^-Z$
               pairs produced per year at LHC ($\sqrt s =~16$ TeV)
               for $M_V=1000~GeV$, $g''=13$ and $z=0$
               within BESS $SU(2)\otimes SU(2)$, with a
               luminosity of $100~fb^{-1}$.
               The applied cuts are: $|y_{W,Z}|<2.5$, $(p_T)_Z>360~GeV$
               and $M_{WZ}>850~GeV$.
               The lower, intermediate and higher histograms
               refer to the background, background plus fusion signal and
               background plus
               fusion signal plus $q\bar q$ annihilation signal, respectively.

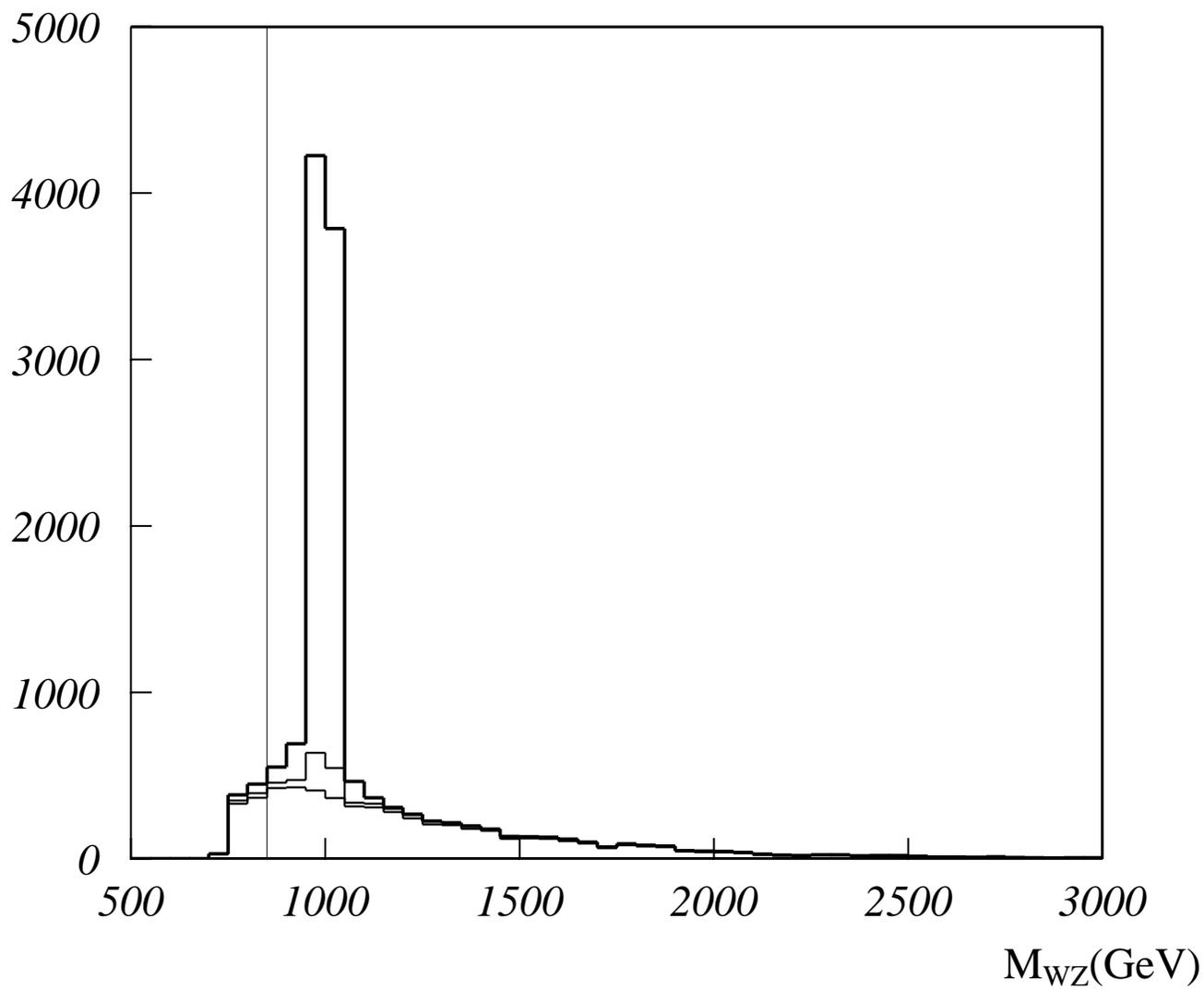
\item [Fig. 9] Invariant mass distribution of the $W^+Z+W^-Z$
               pairs produced per year
               at LHC ($\sqrt s =~16$ TeV) for $M_V=1000~GeV$, $g''=13$
               and $z=0$ within BESS $SU(8)\otimes SU(8)$, with a
               luminosity of $100~fb^{-1}$.
               The applied cuts are: $|y_{W,Z}|<2.5$, $(p_T)_Z>360~GeV$
               and $M_{WZ}>850~GeV$.
               The lower, intermediate and higher histograms
               refer to the background, background plus fusion signal and
               background plus
               fusion signal plus $q\bar q$ annihilation signal, respectively.

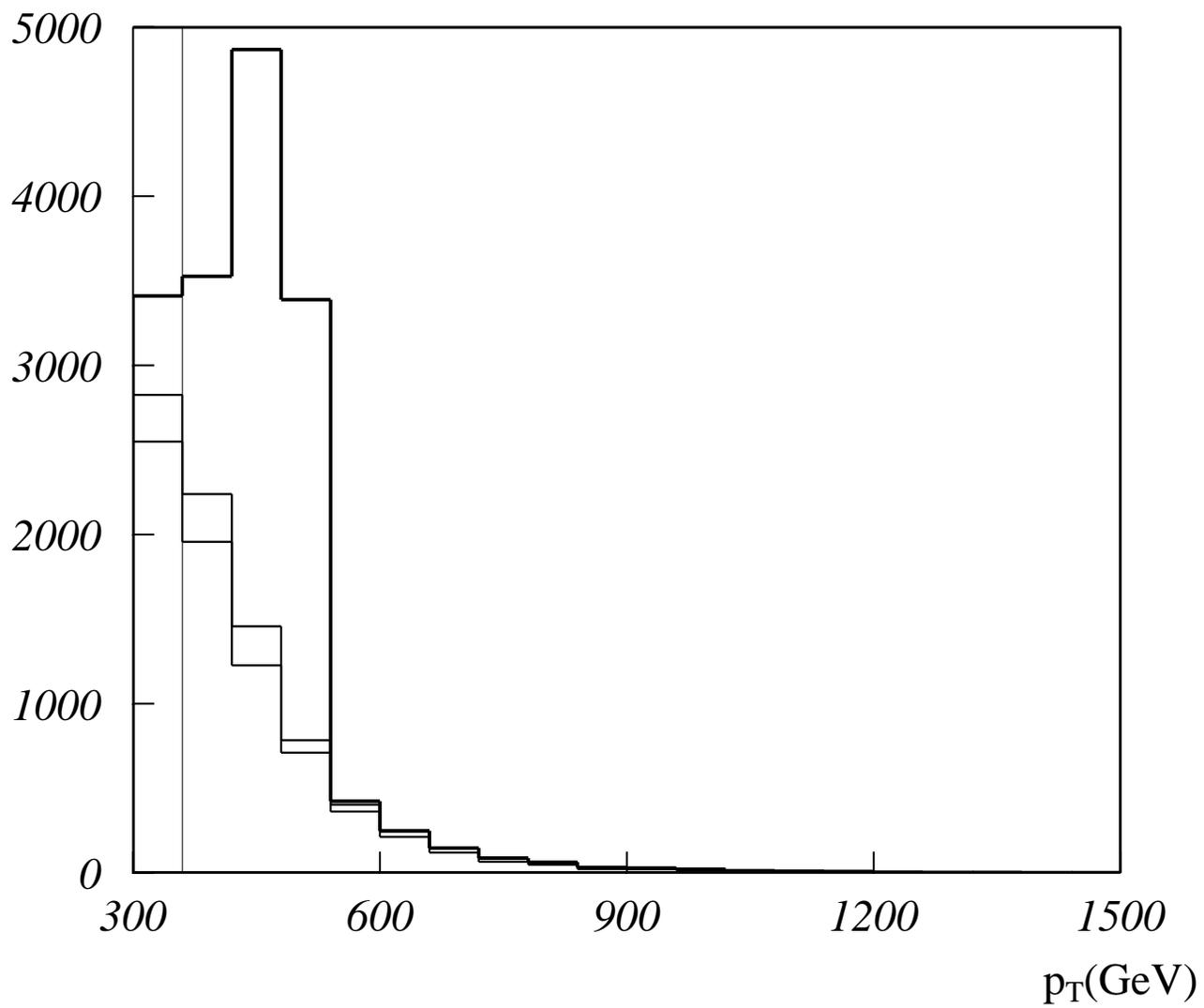
\item [Fig. 10] $(p_T)_Z$ distribution of the $W^+Z+W^-Z$
                pairs produced per year at LHC ($\sqrt s =~16$ TeV)
                for $M_V=1000~GeV$, $g''=13$ and $z=0$,
                within BESS $SU(8)\otimes SU(8)$, with a
                luminosity of $100~fb^{-1}$.
                The applied cuts are: $|y_{W,Z}|<2.5$, $(p_T)_Z>360~GeV$
                and $M_{WZ}>850~GeV$.
                The lower, intermediate and higher histograms
                refer to the background, background plus fusion signal and
                background plus
                fusion signal plus $q\bar q$ annihilation signal, respectively.

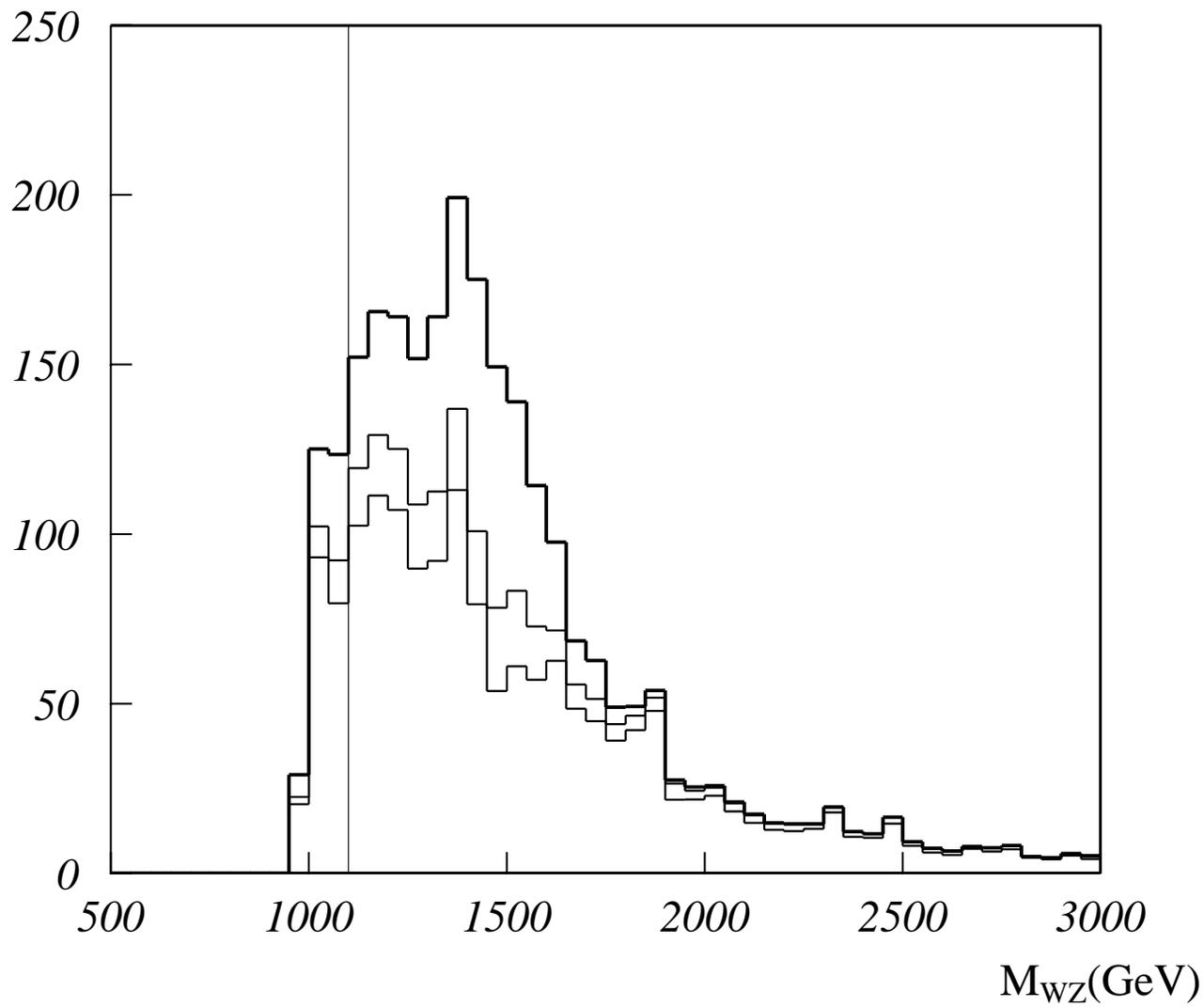
\item [Fig. 11] Invariant mass distribution of the $W^+Z+W^-Z$
                pairs produced per year
                at LHC ($\sqrt s =~16$ TeV) for $M_V=1500~GeV$, $g''=13$
                and $z=0$ within BESS $SU(8)\otimes SU(8)$, with a
                luminosity of $100~fb^{-1}$.
                The applied cuts are: $|y_{W,Z}|<2.5$, $(p_T)_Z>480~GeV$
                and $M_{WZ}>1100~GeV$.
                The lower, intermediate and higher histograms
                refer to the background, background plus fusion signal and
                background plus
                fusion signal plus $q\bar q$ annihilation signal, respectively.

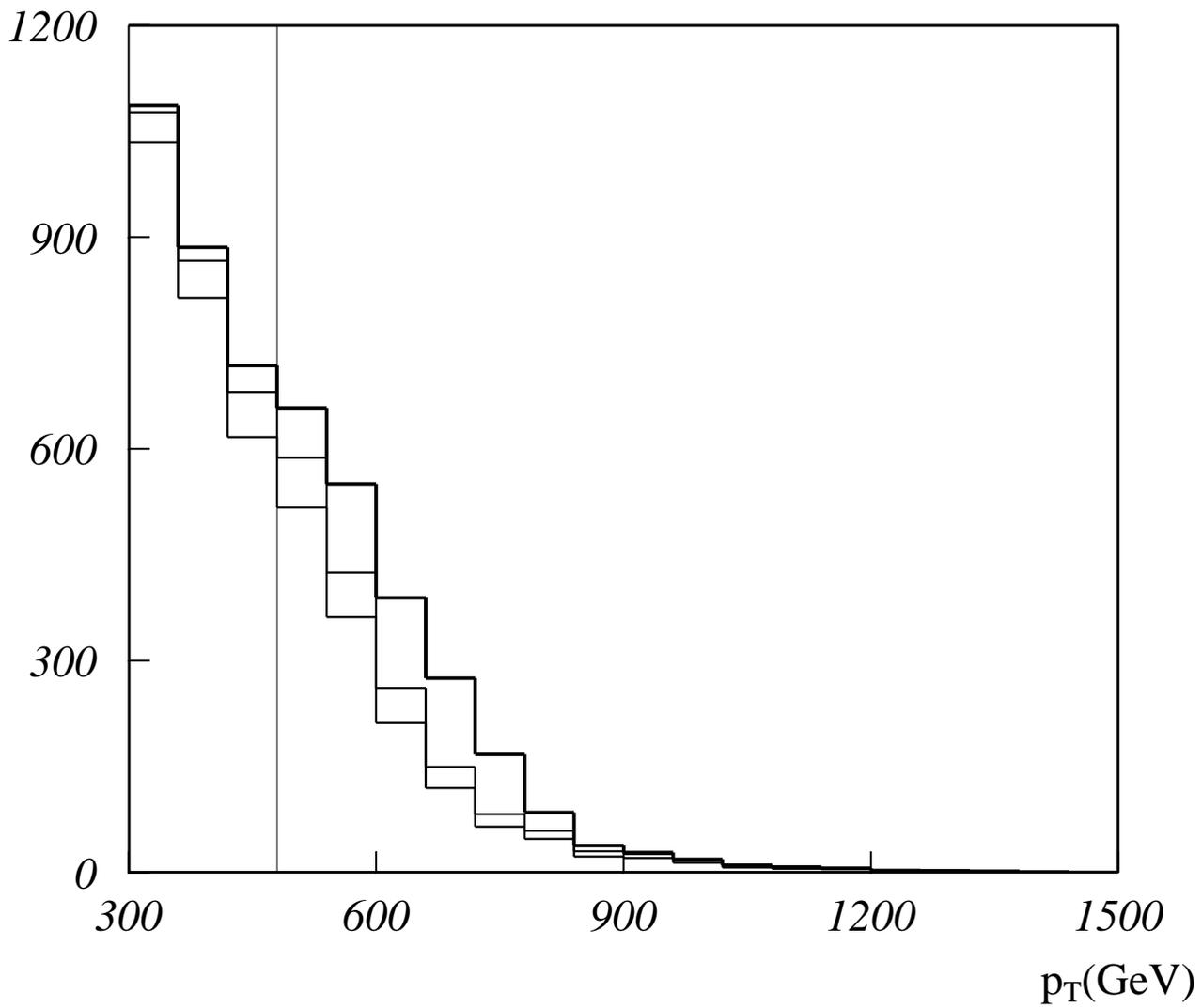
\item [Fig. 12] $(p_T)_Z$ distribution of the $W^+Z+W^-Z$
                pairs produced per year at LHC ($\sqrt s =~16$ TeV)
                for $M_V=1500~GeV$, $g''=13$ and $z=0$,
                within BESS $SU(8)\otimes SU(8)$, with a
                luminosity of $100~fb^{-1}$.
                The applied cuts are: $|y_{W,Z}|<2.5$, $(p_T)_Z>450~GeV$
                and $M_{WZ}>1100~GeV$.
                The lower, intermediate and higher histograms
                refer to the background, background plus fusion signal and
                background plus
                fusion signal plus $q\bar q$ annihilation signal, respectively.

\end{description}
\end{document}